\documentclass[twocolumn, showpacs,preprintnumbers,amsmath,amssymb]{revtex4}

\usepackage{graphicx}
\usepackage{dcolumn}
\usepackage{bm}

\begin{document}

\preprint{APS/123-QED}

\title{On the elliptical flow in asymmetric 
collisions and nuclear equation of state\\}

\author{Varinderjit Kaur}
\author{Suneel Kumar}%
\email{suneel.kumar@thapar.edu}

\affiliation{%
School of Physics and Material Science, Thapar University Patiala-147004, Punjab (India)\\
}%

\date{\today}

\begin{abstract}
We here present the results of elliptical flow for the collision of different asymmetric nuclei 
($_{10}Ne^{20}+_{13}Al^{27}$, $_{18}Ar^{40}+_{21}Sc^{45}$, $_{30}Zn^{64}+_{28}Ni^{58}$, 
$_{36}Kr^{86}+_{41}Nb^{93}$) by using the Quantum Molecular Dynamics (QMD) model. General features of 
elliptical flow are investigated with the help of theoretical simulations. The simulations are performed at
different beam energies between 40 and 105 MeV/nucleon. A significant change can be seen from in-plane to 
out-of-plane elliptical flow of different fragments with incident energy. A comparison with 
experimental data is also made. Further, we predict, for the first time that, elliptical flow for 
different kind of fragments follow power law dependence ${\alpha}$ C${(A_{tot})^\tau}$ for asymmetric
systems. \\ 
\end{abstract}
\maketitle
\section{Introduction}
The heavy-ion collisions at intermediate energies have captured the center place in present day nuclear 
research. This is because of the several rare phenomena emerging at these incident energies and their 
utility in several other branches of the physics \cite{rkpuri94,dasgupta,aichelinRep,khoa}. The nuclear 
equation of state (EOS) is hoped to be pindown via the liquid-gas transition, multi-fragmentation and 
anisotropic flow, such as directed and elliptic flows produced in heavy-ion collisions 
\cite{sood, moretto}. Such knowledge is not only of the interest in nuclear physics but is also useful 
in the understanding of astrophysical phenomena such as the evolution of early universe. One observable 
that has been used extensively for extracting the nuclear EOS from heavy-ion collisions is the anisotropic 
flow of various particles.\\
 In recent years, the subject of anisotropic flow at intermediate energies has attracted increasing 
attention of the heavy-ion community. One of the main reasons for this is that anisotropic flow, 
\cite{moretto, ollitrault} is very 
sensitive to the early evolution of the system \cite{ollitrault,sorge,hartnack}. The anisotropic 
flow is defined 
as the azimuthal asymmetry in the particle distribution with respect to the reaction plane 
(the plane spanned by the beam direction and impact parameter). This gives the magnitude of the asymmetry 
and is characterised by the Fourier expansion of the azimuthal distribution \cite{voloshin} of the detected
particles at mid rapidities as\\
$$\frac{dN}{d\phi} = p_o( 1+2V_{1}Cos{\phi}+2V_{2}Cos2{\phi}).$$\\
Here ${\phi}$ is the azimuthal angle between the transverse momentum of the particle and reaction plane. 
The first harmonic term describes the so-called directed flow which is the bounce-off of cold spectator 
matter in the reaction plane \cite{hstocker80} and the second harmonic term corresponds to the elliptical
flow which is the squeeze out of the hot and compressed participant matter perpendicular to the reaction 
plane \cite{hstocker82,mbtsang,larionov}. The corresponding Fourier coefficients, ${V_n}$ are used to 
quantify this effect \cite{voloshin}. These two observables together represent the anisotropic part of the
transverse flow and appear only in the non-central heavy-ion collisions. The elliptical flow 
\cite{voloshin} has been proven to be one of the most fruitful probes for extracting the equation of state 
(EOS) and to study the dynamics of heavy-ion collisions that originates from the almond shape of the 
overlap zone and is produced due to the unequal pressure gradients i.e an anisotropy in the transverse 
momentum direction. The ${V_2}$ can also be expressed as \\
$${V_2} = {\Large <}\frac{P_x^{2}-P_y^{2}}{P_x^{2}+P_y^{2}}{\Large >},$$~~~~~~ \\
where ${P_x}$ and ${P_y}$ are the components of the momentum along the x and y-axis, respectively.\\
The positive value of ${V_2}$ describes the eccentricity of an ellipse like distribution and indicates the
in-plane enhancement of the particle emission, i.e. a rotational behaviour. Note that isotropy
of the space requires ${V_2}$ should vanish.  
Obviously, zero value of the ${V_2}$ corresponds to an isotropic distribution in the transverse plane. 
Since elliptical flow develops at a very early stage of the nuclear collision, it is an excellent tool to 
probe the nuclear equation of state (EOS) under the extreme conditions of temperature and density.\\
Many theoretical and experimental efforts have been made in studying the
collective transverse flow in heavy-ion collisions at intermediate energies \cite{sorge, calt, yuming, 
westfall}. In 1982, Stocker {\it et al.}, were the first to predict the midrapidity
emission perpendicular to the reaction plane \cite{hstocker82}. At the same time, Gyulassy, Frankel, and 
Stocker \cite{frankel} put forward the kinetic energy tensor to analyze the flow patterns. 
Experimentally observed out-of-plane emission, termed as squeeze-out, was  first observed 
by Diogene collaboration \cite{NPA562}. The Plastic Ball group at the Bevalac in Berkley were the 
first one to quantify the squeeze in symmetric systems. In their study, ${Au+Au}$ collisions was studied 
at 400 MeV/nucleon \cite{gutbrod}. 
Furthermore, the transition from the in-plane to out-of-plane emission was first observed by the NAUTILUS 
collaboration at GANIL in 1994 using the reactions of ${Zn+Ni}$ \cite{popescu}. 
Theoretically, many important develpments in this area e.g the dependencies of elliptical flow on the 
impact parameter and transverse momentum have been determined. These main features have been 
confirmed by the FOPI, INDRA and ALADIN collaborations \cite{lukasik, andronic}. The MINIBALL/ALADIN 
collaboration also observed the onset of out-of plane emission in ${Au+Au}$ collisions at 100 MeV/nucleon 
\cite{mbtsang}. In most of these studies, $_{79}Au^{197}+_{79}Au^{197}$ reactions have been chosen 
\cite{lukasik, andronic}. The elliptical flow at incident energies from tens to hundreds of MeV/nucleon are
 determined by the complex interplay among expansion, rotation and the shadowing of spectators 
\cite{zhang}. Both the mean field and 
two-body collision parts (real and imaginary parts of G-matrix) play an equal important 
role in this energy domain. The mean field plays a 
dominant role at low incident energies, and then
gradually the two-body collision becomes dominant as the incident energy increases. 
It is worth mentioning that the process such as fusion, fission and cluster decay can be
explained successfully using the real part of G-matrix \cite{rkgupta1, rkgupta2}. A detailed study of the
excitation function of elliptical flow, in this energy region, therefore, can provide useful information 
on the nature of nucleon-nucleon interaction related to the equation of state.  Interestingly, a 
systematic theoretical study of the elliptical flow for asymmetric systems is still very rare. 
Moreover, the elliptical flow can also be exploited to understand the nature of hadronic matter 
\cite{modphys}. \\
In this paper, we attempt to study
the different aspects of elliptical flow ${V_2}$ for various asymmetric systems $(A_P \ne A_T)$ e.g., 
$_{10}Ne^{20}+_{13}Al^{27}$, $_{18}Ar^{40}+_{21}Sc^{45}$, $_{30}Zn^{64}+_{28}Ni^{58}$, 
$_{36}Kr^{86}+_{41}Nb^{93}$ within the framework of Quantum Molecular Dynamics (QMD)\cite{aichelinRep} 
Model. This model is used to generate the phase space of nucleons. The article is organized as follows: 
A brief description of the model is given in section II. Our
results are discussed in section III. Finally, we summarize the results in section IV.
\section{The Model}
The quantum molecular dynamics (QMD) model is a time-dependent many-body theoretical approach which is 
based on the molecular dynamics picture that treats nuclear correlations explicitly. The two dynamical 
ingredients of the model are the density-dependent mean field and the in-medium nucleon-nucleon cross-
section\cite{bohnet}. In the QMD model, each nucleon is represented by a Gaussian wave packet characterised
 by the time dependent parameters in space ${\vec{r_i}(t)}$ and in momentum ${\vec{p_i}(t)}$ 
\cite{aichelinRep}. This wave packet is represented as\\
\begin{equation}
\Phi_i(\vec{r},\vec{p},t) = \frac{1}{(2{\pi}L)^{3/4}}e^{\frac{{-(\vec{r}-\vec{r}_i(t))^2}}{2L}}e^{\frac{i\vec{p}_i(t).\vec{r}}{\hbar}}.
\end{equation}
The total n-body wave function is assumed to be the direct product of form\\
\begin{equation}
\psi = \prod_{i=1}^{A_T+A_P} \Phi_i.
\end{equation}
The model uses its classical analog in terms of the Wigner function \cite{carruthers}. 
\begin{equation}
f_i(\vec{r},\vec{p},t) = \frac{1}{({\pi}{\hbar})^{3}}e^{\frac{{-[\vec{r}-\vec{r}_i(t)]^2}}{4L}}e^{\frac{{-[\vec{p}-\vec{p}_i(t)]^2}.2L}{\hbar^2}.}
\end{equation}
The parameter L is related to the extension of the wave packet in phase space. This parameter is, however, 
kept fixed in the present study. The centroid of each nucleon propagates under the classical Hamilton 
equations of motion \cite{aichelinRep}.\\


\begin{equation}
\frac{d\vec{r_i}}{dt} = \frac{dH}{d\vec{p_i}}~~~~;~~~~\frac{d\vec{p_i}}{dt} = -\frac{dH}{d\vec{r_i}}~.
\end{equation}
Here H, the Hamiltonian, reads as ${ <H> = <T> + <V>}$. The local potential has the break up as:  

\begin{equation}
V^{tot} = V^{Loc} + V^{Yukawa} + V^{Coul} + V^{MDI}. 
\end{equation}
Here ${V^{Loc}}$,  ${V^{Yukawa}}$,  ${V^{Coul}}$, and  ${V^{MDI}}$ represent, respectively, the Skyrme, Yukawa, Coulomb, and Momentum-dependent (MDI) parts of the interaction. The ${V_{Loc}}$ and ${V_{Yukawa}}$ 
further reads as: \\
\begin{equation}
V^{Loc}=t_1\delta(\vec{r_1}-\vec{r_2})+ t_2\delta(\vec{r_1}-\vec{r_2})\delta(\vec{r_1}-\vec{r_3})~,
\end{equation}
and\\
\begin{equation}
V^{Yukawa}=t_3\frac{exp({\mid{\vec{r_1}-\vec{r_2}}\mid}/m)}{({\mid{\vec{r_1}-\vec{r_2}}\mid}/m)}~,
\end{equation}
with $m$ = 1.5 fm and $t_3$ = -6.66 MeV.\\

The static (local) interaction can be parametrized as\\
\begin{equation}
U^{loc} = \alpha (\frac{\rho}{\rho_o})+ \beta (\frac{\rho}{\rho_o})^2.
\end{equation}
The above two parameters ${(\alpha, \beta)}$ are fixed by the requirement that the average binding energy
(at normal nuclear matter density ${\rho_o}$) should be -15.76 MeV and the total energy should have a 
minimum at ${\rho_o}$. In order to understand the role of different compressibilities, the above 
potential can be generalized to\\
\begin{equation}
U^{loc} = \alpha (\frac{\rho}{\rho_o})+ \beta (\frac{\rho}{\rho_o})^{\gamma}.
\end{equation}
The momentum dependent interaction is obtained by parametrizing the momentum dependence of the 
real part of optical potential. The final form of the potential reads 
as \cite{aichelinRep, sanjeevvermani}\\
\begin{equation}
U^{MDI} = t_4{\ln}^2[t_5(\vec{p_1}-\vec{p_2})^2+1]\delta(\vec{r_1}-\vec{r_2}).
\end{equation}
Here ${t_4}$ = ${1.57~{MeV}}$ and  ${t_5 = 5 \times{10}^{-4} {MeV}^{-2}}$. A parametrized form of 
the local plus momentum dependent interaction (MDI) potential (at zero temperature) is given by \\
\begin{equation}
U = \alpha (\frac{\rho}{\rho_o})+ \beta (\frac{\rho}{\rho_o})^{\gamma}+ \delta{\ln}^2
[\epsilon(\frac{\rho}{\rho_o})^{\frac{2}{3}} + 1]\frac{\rho}{\rho_o}.
\end{equation}
The different parameters appearing in Eq. (11) are taken from the Ref. \cite{suneel}.\\

The different values of the compressibility in the Skyrme force give possibility to look for the role of 
different equations of state termed as soft and hard equations of state. The inclusion of momentum 
dependent interactions in soft and hard equations of state are labeled as soft momentum dependent (SMD) EOS
 and hard momentum dependent (HMD) EOS, respectively. The (additional) momentum dependence of the 
interaction generates extra repulsion during the 
evolution of the reaction which increases the collective flow and supresses the subthreshhold particle 
production \cite{huang}. This effect is the largest during the initial phase of the reaction. The 
imaginary part of the potential is parametrized in terms of the nucleon-nucleon cross-section. As our 
present aim is to look for the role of different EOS's, we use the energy dependent NN cross-section. A 
relativistic version of the model is also available in the literature \cite{lehmann}.\\

\section{Results and Discussions}
\label{sec:2}
The main advantage of the QMD model is that it can explicitly represent the many body states of the
system and thus contains the correlation effects. Therefore, provides an important information about the 
collision dynamics. \\

In this paper, the free nucleons and 
different fragments i.e., light mass fragments (LMF's)${(2\le A \le4)}$ and intermediate mass fragments 
(IMF's)${(5\le A \le{A_{tot}}/6)}$ are selected to study the elliptical flow analysis. Some studies 
have been carried out for elliptical flow involving heavier fragments \cite{1994}. We perform a complete 
systematic 
study using the different asymmetric  reactions  $_{10}Ne^{20}+_{13}Al^{27}$, $_{18}Ar^{40}+_{21}Sc^{45}$, 
$_{30}Zn^{64}+_{28}Ni^{58}$, $_{36}Kr^{86}+_{41}Nb^{93}$ at incident energies between 45 and 105 
MeV/nucleon. We have simulated these reactions over the full collision geometry starting from the central 
to peripheral one. The fragments are constructed within minimum spanning tree (MST) method 
\cite{aichelinRep}. The MST method binds two nucleons in a fragment if their distance is less than 4 fm. 
The entire calculations are performed at final state that is 200 fm/c. We also note that several different 
models on clusterisation are also reported in recent publications \cite{vermani}.\\ 
In fig.1, the final state elliptical flow is displayed for the free particles (upper panel), light charged 
particles (LCP's)${(2\le A \le4)}$(middle),
 and intermediate mass fragments (IMF's)${(5\le A \le{A}_{tot}/6)}$ (lower panel) 
at the scaled impact parameter  $\hat{b}$ = 0.5. One can see a Gaussian type behaviour 
quite similar to the one reported by Colona and Toro {\it et al.},
 \cite{colona}. This Gaussian type behaviour is integrated over the entire rapidity range. One also sees 
from the figure, that the elliptical flow of nucleons/LMF's/IMF's is positive in the whole range of 
${P_t}$. There is a linear increase in the elliptical flow with transverse momentum upto certain incident 
energy. Obviously, particles with larger momentum will escape the reaction zone earlier. After certain 
transverse momentum, it starts decreasing monotonically, which is due to the decrease in the number of 
particles in that ${P_t}$ range. One also notices that the peaks of Gaussian shifts toward the lower value 
of ${P_t}$ for heavier fragments. The reason is that the emitted free and light charged particles can feel 
the role of mean-field directly, while the heavier fragments have weak sensitivity \cite{yan}. It is also 
evident from the figure that the reactions under consideration have different system mass dependence as 
well as N/Z ratios.\\ 

Moreover, the results are found to vary drastically with the change of equation of state.
To make this point more clear, we have dislayed on the right hand side of the fig.1, the transverse 
momentum dependence of elliptical flow using four different equations of state namely the soft, hard, 
soft momentum dependent (SMD), and hard momentum dependent (HMD) equations of state. For this analysis, we 
choose the reaction of Ar+Sc at $\hat{b}$ = 0.5 and E = 95 MeV/nucleon. We can see clear effect of the 
different equations of state in the production 
of light mass fragments (LMF's). In contrary, very little change is observed in the case of free particles 
and intermediate mass fragments (IMF's). This is due to the different compressibility values for hard, HMD 
EOS (K= 380 MeV), and soft, SMD EOS (K= 200 MeV). Moreover, the role of momentum dependent 
interactions (MDI) is also important. Because of the repulsive nature of momentum dependent 
interactions, which leads to the suppression of binary collisions, less squeeze-out is observed in 
the presence of momentum dependent interactions (SMD, HMD) compared to the static one (S, H) 
\cite{andronic}. Since
the figure is plotted at all rapidities, so the less positive value of elliptical flow is obtained
in the presence of momentum dependent interactions. Therefore, here we investigate the influence of 
different EOS's on the elliptical flow related to the fragments having different sizes. \\ 
  
\begin{figure}
\includegraphics {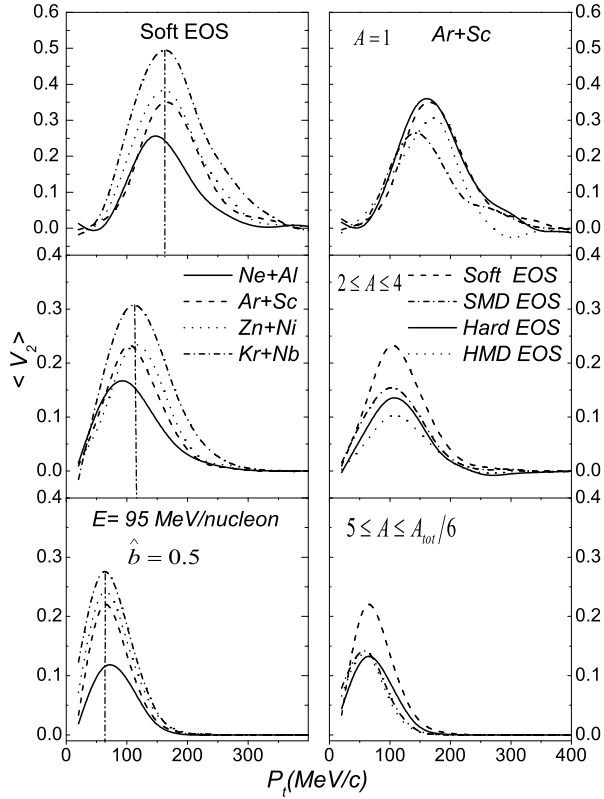}
\caption{\label{fig:1} The elliptical flow as a function of the transverse momentum at incident
energy 95 MeV/nucl. for free particles, LMF's and IMF's in top, middle and bottom panels respectively. 
On the right hand side, elliptical flow for different EOS's for Ar+Sc system has been displayed.}  
 \end{figure} 
\begin{figure}
\includegraphics {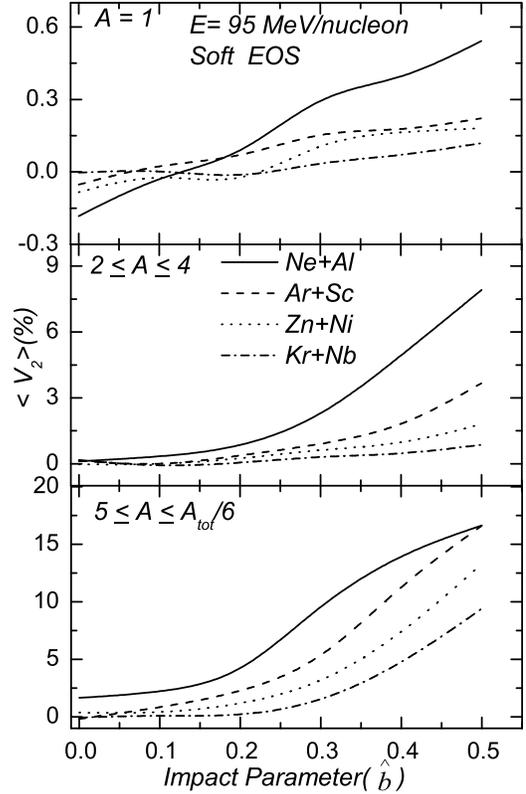}
\caption{\label{fig:2} The elliptical flow as a function of impact parameter and at incident 
energy  E = 95 MeV/nucl. for free particles, LMF's and IMF's in top, middle and bottom panels 
respectively.}  
\end{figure}
The investigation of the elliptical flow with scaled impact parameter for different asymmetric systems is 
displayed in fig.2. The simulations have been carried out at 95 MeV/nucleon for the scaled impact 
parameter range between $\hat{b}$ = 0.0 and 0.5. The results for the free nucleons (upper panel), LCP's 
(middle panel) and IMF's (lower panel) have been shown for all four systems considered in this analysis. 
The value of the elliptical flow ${V_2}$ increases  with the impact parameter. There is an obvious 
difference among the elliptical flow associated with the free particles(upper panel), light charged 
particles (LCP's)${(2\le A \le4)}$(middle), and intermediate mass fragments (IMF's)
${(5\le A \le{A}_{tot}/6)}$(lower panel). This is more true at higher impact parameters. One can see the 
isotropic distribution of the fragments at small impact parameters which turns into anisotropic 
distribution as we move away towards semi-central zone. For lighter 
systems, the value of ${V_2}$ increases more rapidly with impact parameter. This is due to the azimuthal 
anisotropy, that becomes larger with the impact parameter and reduces with the beam energy. For semi-
central collisions, less number of nucleons participate in the collision process and hence lead to more 
elliptical flow. Since the production of the intermediate mass fragments is due to the spectator part, the 
elliptical flow is stronger for the heavier fragments compared to the lighter colliding nuclei. \\

\begin{figure}
\includegraphics {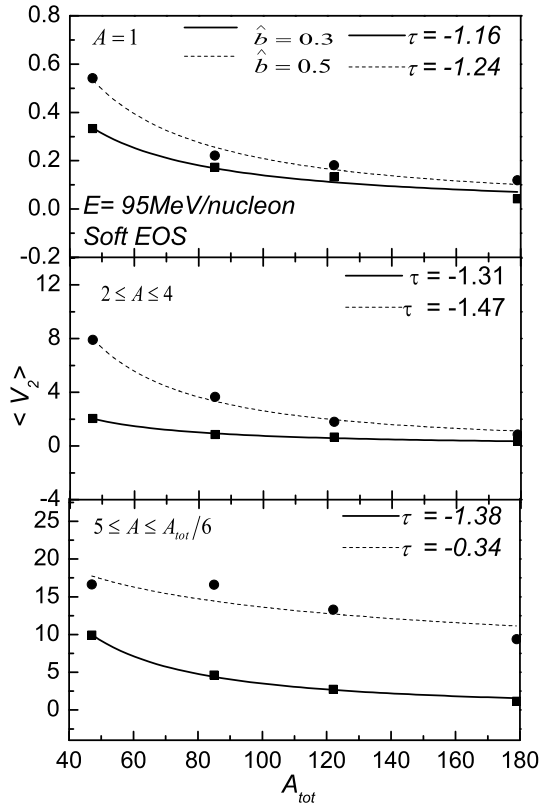}
\caption{\label{fig:3} The dependence of elliptical flow on the composite mass of the systems for free
nucleons, LMF's and IMF's in top, middle and bottom panels, respectively. The curves are fitted with a 
power law behaviour ${\alpha}$ C${(A_{tot})^\tau}$.}  
\end{figure}
We further carry out the study of the system size dependence of the elliptical flow of free particles 
(upper panel), light charged particles (LCP's)${(2\le A \le4)}$(middle), and intermediate mass fragments 
(IMF's) ${(5\le A \le{A}_{tot}/6)}$ in different asymmetric reactions, namely $_{10}Ne^{20}+_{13}Al^{27}$, 
$_{18}Ar^{40}+_{21}Sc^{45}$, $_{30}Zn^{64}+_{28}Ni^{58}$, $_{36}Kr^{86}+_{41}Nb^{93}$ at two different 
impact parameters $\hat{b}$ = 0.3 and $\hat{b}$ = 0.5. The elliptical flow varies with the mass number. 
This is not surprising. It was shown that there is a linear relation between the system size and particle 
emission \cite{jsingh,gutbrod}. 
One can see that elliptical flow decreases with composite mass of the system, 
while the reverse trend is observed with the impact parameter. The reason for this is that the pressure 
produced by the Coulomb interactions increases with the system size, while, decreases with the impact 
parameter. The 
decrease with impact parameter is due to the decrease of the participant zone. All the curves are fitted 
with a power law behaviour ${\alpha}$ C${(A_{tot})^\tau}$, where
C and ${\tau}$ are the fitting parameters.\\

\begin{figure}
\includegraphics{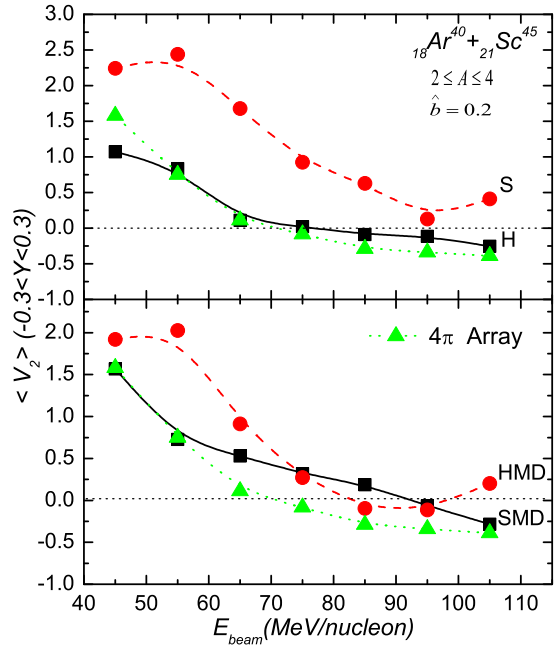}
\caption{\label{fig:4}(color online) The variation of the elliptical flow with beam energy for the reaction of 
${Ar+Sc}$ using hard, soft, HMD, and SMD equations of state. Here comparison is shown with 
experimental findings of the 4${\pi}$ Array \cite{magestro}.} 
\end{figure}
In fig.4, we show the elliptical flow ${V_2}$  for the midrapidity region ${(-0.3\le Y \le0.3)}$ for 
${2\le A \le4}$ as a function of the incident energy. The rapidity cut is in accordance with the 
experimental findings. The theoretical results
are compared with the experimental data extracted by the ${4\pi}$ Array group \cite{magestro}. For the 
comparison of calculations with experimental findings, we have performed the detailed analysis by taking 
into account all equations of state. The comparison has been done at the same impact parameter as suggested
 by Magestro \cite{westfall, magestro, rpak} after
correction. The general behavior of the elliptical flow 
calculated with different EOS's resembles. The elliptical flow evolves from a preferential in-plane 
(rotational like) emission (${V_2 > 0}$), to the out-of-plane (squeeze-out) emission (${V_2 < 0}$), with 
an increase of the incident energies. The elliptical flow decreases with the incident energy. The decrease 
in elliptical flow is sharp at smaller incident energies (upto 65 MeV/nucleon). This starts saturating at 
higher incident energies. In other words, the elliptical shape of nuclear flow dominates the physics at low
 incident energies. One can see from the figure that the transition energies at which the elliptical flow 
parameter ${V_2}$ changes sign from positive to negative are different for different equations of state. 
The transition from the in-plane emission to out-of plane occurs because the mean-field that contributes 
to the formation of the compound nucleus becomes less 
important. At the same time, the collective expansion process (based on the nucleon-nucleon scattering) 
starts dominating. The competition between the mean-field and the nucleon-nucleon collisions strongly 
depends on the effective interaction , which leads to the divergence of the transition energies calculated 
with different equations of state. Clearly the hard EOS provides stronger pressure that leads to a stronger
 out-of plane emission and thus to the smaller transition energy.  The difference of elliptical flow 
between all EOS's 
decreases as we move towards the higher incident energies (E = 85 to 105 MeV/nucleon). The elliptical 
flow calculated with hard equation of state agrees with the ${4\pi}$ Array experimental 
data. From our analysis, we find that hard equation of state explains the data closely. 
We predict, for the first time 
that, elliptical flow for different kind of fragments follow power law dependence 
${\alpha}$ C${(A_{tot})^\tau}$ for asymmetric systems.  \\ 

\section {Conclusion}
We have presented systematic theoretical results on elliptical flow by analyzing 
various asymmetric reactions e.g., $_{10}Ne^{20}+_{13}Al^{27}$, $_{18}Ar^{40}+_{21}Sc^{45}$, 
$_{30}Zn^{64}+_{28}Ni^{58}$, $_{36}Kr^{86}+_{41}Nb^{93}$ at incident energies between 45 and 
105 MeV/nucleon and over full geometrical overlap in the framework of the QMD model. The general features 
of the elliptical flow are investigated with the help of theoretical simulations, particularly, the 
transverse momentum, impact 
parameter, system size dependence and incident energy dependence. A special emphasis was put on the energy
dependence of the elliptical flow. We also compared our theoretical calculations with ${4\pi}$ 
Array data. This comparison, performed for the reaction of $_{18}Ar^{40}+_{21}Sc^{45}$ shows that the
elliptical flow is sensitive to all the equations of state that have been used i.e., hard,
soft, HMD, and SMD equations of state. In general, a good agreement is obtained between the data 
and calculations.
We predict, for the first time 
that, elliptical flow for different kind of fragments follow power law dependence 
${\alpha}$ C${(A_{tot})^\tau}$ for asymmetric systems. 
\section {Acknowledgment}
This work has been supported by the grant from Department of Science and Technology (DST) Government of 
India vide Grant No.SR/WOS-A/PS-10/2008.\\ 

\end{document}